# SPECTROSCOPIC MANIFESTATION OF DYE PAIR INTERACTIONS AT HIGH CONCENTRATIONS IN STRUCTURALLY ORGANIZED $SiO_2$ FILMS

Short version of title: **SPECTRUMS OF DYES IN THE MESOPOROUS SILICATE FILM**

**E.A.Tikhonov**, Physical institute NAS Ukraine, Kiev, job #: 38(044) 5250592, a fax #: 38(044) 5251589, E-mail: etikh@iop.kiev.ua
**G.M.Telbiz,** Physical chemistry institute NAS Ukraine, Kiev

## ABSTRACT

Concentration spectral manifestation of the ionic and neutral dyes encapsulated in mesoporous silicate films was studied. The mesoporous dyed $SiO_2$ film was prepared by sol-gel technology with use as surfactant the amphiphilic three-bloc copolymer Pluronic P123. It was discovered when dye concentration exceeds mesoporous one the new spectral bands appear that are distinct from the initial monomolecular predecessors. To explain the phenomena guess is formulated and proved that origin of new spectrums is obliged to formation of the ionic pairs, excimers and excitons with participation of dye molecules localized in the characteristic places of mesoporous silicate film.

## INTRODUCTION

Organic dyes have occupied a long time ago extensive spheres of applications, nevertheless remained as objects of incessant scientific attention up to nowadays. In particular, application of dyes in the laser physics and optics /1/ have boosted the development sol-gel of technology with the purpose manufacturing of organically modified silicates (ormosil) having a priori the best the thermo-optical parameters in comparison with polymeric matrixes /2/. In research work /3/ it has been found that photochemical stability of covalently bound dyes in ormosil is higher in compare to case of Van-der-Waals bonds in solutions.

Other technology of organic dye encapsulation in a glass matrix has originated in connection with production engineering of the porous and mesoporous silicate glasses. The glasses can be distinguished with the accidental and regular organized pores. For manufacturing of the similar porous matrixes it has been developed methods of acid-alkaline etching of soda-borate-silicate glasses /4/. Dye embedding in the similar matrix is carried out by sorption the dye from the solution. Because the ionic dyes get the necessary spectroscopic parameters only in the polar solvents, pores of such matrix should be filled with the corresponding solvent. Laser application the similar dyed mesoporous structures for light oscillation have been carried out in research /5/.

For the first time the regular mesopores in silicate glasses prepared by sol-gel technology have been gained by authors of contribution /6, 7/. However application with the specified purpose not ionic di - and three-bloc copolymers – surfactants of new generation has been begun from research /8/. The ability of amphiphilic molecules to self-assembly in water solutions in modular structures - micelles - with the subsequent self-organizing of micelles in spatially ranked ensembles of various symmetry plays the dominant role in the given approach. If in the first stages of the approach it was known only some simple ionic surfactants (ammonix LO) capable to raise a dye solulability and thereby to interfere their dimerization to conserve the fluorescence quantum yield, for today their assortment is quite wide and is characterized by the extensive actual and perspective technical/medical applications. Well-defined block copolymer surfactants undergo self-assembly in aqueous solution in order to minimize energetically unfavorable hydrophobic interactions. In accordance with the theoretical model /9/ specific self-assembled nanostructures can be targeted according to a dimensionless «packing parameter» p, which is defined in following equation: **p=v/al**, where **v** is the volume of the hydrophobic chains, **a** is the optimal area of the head group, and **l** is the length of the hydrophobic tail. Therefore, the packing parameter "p" of a given amphiphilic molecule usually dictates its most likely self-assembled morphology.

As a general rule spherical micelles are favored when $p < 1/3$, cylindrical micelles when $1/3 < p < 1/2$, and enclosed membrane structures (vesicles, also known as polymersomes) when $1/2 < p < 1$.

## STATEMENT AND MOTIVATION OF INVESTIGATION

Studying of spectroscopic behavior of ionic, neutral and zwitterionic dyes (on one representative from the termed group) at extremely high achievable concentrations in structurally ranked silicate films (SRSF) was the purpose of the given research. SRSF nanometrical thickness were prepared by the sol-gel technology with use the surfactant Pluronic P123.The hydrophobic nature of the surfactant can be utilized for the strong solubilization of organic compounds and the one-dimensionality enables us to design a well-aligned molecular system inside the hydrophobic nanochannels. In the literature it was cited some spectral changes in fluorescence bands of Rhodamine 6G dye on the high concentration in structured mesoporous silicate films, however without any explanation of observable effect /10/. The most studied and understood feature for similar materials there is an enhancement of electronic energy transfer between donor-acceptor pair in accordance with Forster model at localisation of the molecular pair in micelle or nearby /11, 12/. By and large our statement of questions in the given research is similar to that in to paper /12/ with a such distinctiveness that authors /12/ as surfactant used the ionic polymeric compound cetyltrimethylammonium bromide (CTAB), while we have used Pluronix P123.

## PREPARATION AND TESTING OF SAMPLES

The mesoporous $SiO_2$-films were gained with sol-gel synthesis at presence nonionic three-bloc copolymer as surfactant: $PEO_{20}PPO_{70}$ $PEO_{20}$ (Pluronic P123, Sigma Aldrich, m.w. 5800). In the used technology two separately prepared solutions are mixed and deposited on substrates at their gyration (spin-coating) or drive (dip-coating) with the subsequent drying at the constant room temperature and dampness. Sol-solution was plotted on the base of chemically pure tetraetoxy silane ($Si(OC_2H_5)_4 \equiv$ TEOS) , ethanol, twice distilled water and 35% HCl. After hydrolysis and sol production to this solution it was added earlier prepared gel- solution of the termed dyed surfactant Pluronic P123 at the first critical concentration that caused the spatial self-assembling of already existing micelles. Relation of concentrations of gel-intermixtures (dye and Pluronic P123) in gram-molecule units varied from 1:32 to 32:1. The basic regular composition of the material was described with following relation of molar concentrations: TEOS:$C_2H_5OH$: HCL:$H_2O$=1:20:0.5:8. The necessary aging degree of the terminated intermixture was reached during (40÷60min) at the continuous stirring. Then by mentioned above actions the film deposition at gyration/drive velocities 2040 rpm and 0.1 - 20 sm/mines was carried out on the spacially purified glass substrates (or others materials). The gained liquid films dried at the regular temperature/damp forming the firm silicate coating in the thickness 100-250nm on the used substrate. Owing to evaporation of liquid water fraction in time of drying the concentration of dyed micelles in film increased to second critical value due to that their spatial streamlining takes place. Simultaneously formation occurs the porous grid structure $SiO_2$ with a wall thickness s (3÷5)nm that were built around spatially ordered dyed micelle structure. For the specified small thickness of films streamlining of micelles occur simultaneously in planes of a film and on thickness with formation of cylindrical columns normally directed to the developed film planes. At the intended conservation of the organic micelles in gained film this silicate structure is termed the mesoporous one. Heating of samples to $500^0$C is accompanied by evaporation the organic micelles and formation of the truly porous silicate films.

Formation of the ranked structure in films is tested by methods of the small angular X-ray diffraction and the gained results of testing are presented on fig.1. Results are gained with diffractmeter DRON-3M. They confirm formation of the ordered structure and its structure invariance at use of nonamphiphilic dyes.

The IR spectroscopy specified in almost full elimination of water from the ready films and measured by methods of atomic force microscopy a film thickness was found systematically in limits 180-220nm at an invariance of embodying of the featured technology.

The following evident estimates allows to determine the key parametres of self-assembly spatial structures**.** The wave length of the X-ray radiation for tube with the copper cathode is equal $\lambda_{CuK\alpha}$=1,5405$A^0$. Requirement of Bragg conditions: $m\lambda=2\Lambda\sin(\Theta_m)$ and angular positions of diffraction peaks lead to spatial periods of 2D structures for m=1,2 $\Lambda_1$=102,04$A^0$ and $\Lambda_2$=101,55$A^0$ correspondently. Additional registered maximum on fig.1. is possibly related to hexagonal symmetry of micelle self-organisition. Also it is seen that introduction of ionic dye R6G in gel fraction did not destroy the arising ordered $SiO_2$ structure.

Because ordering and spatial period of the ordered $SiO_2$ film structure is spotted by the spherical micelle sizes owing to the dense packing the micelle concentration at cubic symmetry is determined equal ≈$\mathbf{10^{18}sm^{-3}}$. For the more dense hexagonal packing one can consider concentration of micelles approximately on 1 order higher. From the data fig. 2. on optical density magnitude $\mathbf{D=c\varepsilon T}$=1,5 for R6G with its molar extinction $\varepsilon_{max}$=118*$10^3$ and the film thickness T=200nm it's possible to calculate the dye concentration =0,6m/L=3,6*$10^{20}cm^{-3}$. The comparision of the micelles and dye concentrations by data fig. 2. allows to conclude that 10 times excess concentrations of dye above micelles still does not lead to optically registered interactions between dye molecules in the basic state, which would be felt by the optical absorption spectroscopy.

Prior conclusion from the gained concentration relation between of dye and micelles can be formulated as the following guess: all dissolved dye in gel is not implanted micelles of a silicate film. Its part exceeding the limiting dissolvent power of micelles P123 can be placed in the thermodynamically labile interface region between silicate walls and micelles. It is most likely therefore the initially homogeneous dye distribution in film breaks due to the diffusion dye redistribution with the positive gradient to periphery, especially appreciable on samples at concentrations exceeding the limiting value of 0,5m/l.

## RESULTS AND DISCUSSION.

Subsequent results trace changes in absorption/fluorescence spectrums of the following dyes – ionic R6G, neutral- 6-aminophenalemen and zwitterionic - pyrromethane PM580 with concentration growth in SRSF and gel solution P123 at room temperatures. The used method of analysis promotes to the qualitative understanding of the nature the registered changes in interrelation with the probable localization of ions and the uncharged molecules in SRSF. On fig. 3. evolution of absorption spectrum R6G is presented: curve 1 corresponds to the upper concentration limit, at which the monomolecular fraction exist preferentially; curves 2,3,4 show changes at excess of concentrations 0,5m/l. Changes develop in the weight increase of curve with peak on a wavelength 510nm concerning the standard peak of R6G in polar solvent 535nm.

Observed behavior of R6G spectrum do not suggest to consider the absorption peak 510nm as the vibronic oscillations of $\pi$- electron conjugation chain, as it is accepted habitually. It is more likely to connect the peak with specificity of charge states of ionic dyes, in particular, a cation-anion pair R6G, proving at conditions of incomplete ionisations on ions, in opposite to the typical state R6G in polar solvents at the full ionization on two ions. With application the mathematical procedure of deconvolution the gained integrated contours expands into 2 Gaussian components whose weight contribution changes with concentration R6G but the wavelength peak position conserves. The long-wave absorption band we carry to known cation R6G absorption spectrum, meanwhile a short-wave band is reasonably to connect with ionic pair. The weight contribution of this center depends from dye concentration. The mathematical analysis has been spent in support of the guess about 2 central nature of the absorption spectrum R6G at concentrations above 0,5m/L in SRSF with used technology of preparation.

The behavior of fluorescence spectrum R6G at change its concentrations in SRSF specifies on the multicentre character also. The presented on fig.4a. fluorescence spectrum R6G in SRSF possess the series unusual features. Contour 1 is related with typical cationic spectrum R6G in polar solvents. The contour asperities on its long-wave wing which are absent usually in a cation fluorescence band the dye in polar solutions here appear already at the critical concentrations ($\sim$0,5m/l). It is also impossible to carry them as display of vibronic interactions of $\pi -$ electronic oscillator with vibrations of the molecular skeleton in an excited electronic state: the concentration dependence of the weight contribution of these contours testifies about their multicentre origin.

The deconvolution of integrated fluorescent contour on Gaussian components is presented on fig. 4b. It is visible also that at some excess of critical dye concentration the main contour of cation R6G fluorescence disappears completely and there is only feeble hint on its former existence (the mark on the left wing of the contour 2).

Fluorescence quantum yield R6G in SRSF at these concentrations (always heightened in comparison with usual solutions) multiply decreases, but remained fluorescence power is easy to register by standard technique even at the further concentrations growth up to the polycrystalline powder. Fluorescence spectra were registered by the spectrometer (Hitachi MPF-4)) under condition of the excitation on 530nm that corresponds to the absorption peak of cation R6G and in the light reflection configuration to diminish the reabsorption shifts.

So, the fluorescence band 2 with peak on 605nm grows from the first inflexion on curve 1 (it is marked by arrow). The band contour does not change any more at successive concentration growth, but the concentration increase stipulates the grows of 2nd fluorescence band with peak on 640nm from the second marked inflexion on the initiating fluorescence band.

Registered with more sensitive spectrometer R6G emission spectrums at the highest concentrations are presented on fig.5. Here the fluorescence spectrum of polycrystalline R6G (5) is presented also. Coinciding in relation of the red shift direction with spectrums R6G for the highest concentrations in SRSF, this crystalline R6G spectrum has considerably narrow spectral width. The last specifies the higher degree of streamlining and homogeneity of fluorescence centers (dye molecules) in the case of natural crystal in comparison with quasycrystall streamlining in SRSF.

The multicenter character of fluorescence spectrums proves to be true also from measuring of lifetimes within an integrated contour 3 fig.5. On fig.6a, b, c, d measured data with use for fluorescent excitation the laser radiations of picosecond duration on 357nm are presented. The fluorescent lifetime of the polycrystalline phase R6G was equal 2,7ns. Lifetimes of monomolecular cation shape R6G and a fluorescent center with a contour peak on 605nm (a contour 2 and 3 on fig. 5.) in SRSF occurred equal 2,8ns and 3,5ns that is explainable as the presence of spectrally imposed 2 bands from two centers of the various nature. Lifetimes of fluorescent centers related to a spectrum 4 on fig.5. for wavelength 640nm and 670nm occurs to be equal 2,2 and 2,5нс.

The obtained data allows to draw a deduction about multicenter origin of fluorescence registered at highest concentrations R6G in SRSF above 0,5m/l.

The similar featured for R6G changes in absorption/fluorescence spectrums were observed for zwitterionic dye PM580 (fig.7.), which intermolecular ionization and, correspondently, the spectral position absorption bands depends on polarity of an environment or, other words, from the places of molecular localization in SRSF.

Meanwhile in gel solution P123 the form-factor of absorption band R6G, PM580 did not change in a wide interval of concentrations and coincided with contours 1, 5 on fig. 2, 7 for low consentrations. The neutral for charge aminophenalemine discovered no changes in the form of initial absorption bands in wide limits of dye concentration in SRSF.

A spectral nonhomogeneous broadening in the considerable case is not simply related to casual character of a local field similar to case of rare-earth element ions in the common glass matrix. At the typical case of the nonuniform spectral broadening the continuous shift of a transition frequency for the base centre takes place. In the case under consideration there is a formation of 2-3 centers with the own resonant transitions frequencies. Main reason occurrence of several centers consists in some various possibilities of dye localization in the structurally ranked silicate due to big polymeric micelle ordering in film and the subsequent pair interactions of dye ions (molecules) in the ground and excited states. Therefore let us discuss possible places of localization of the studied dyes in SRSF.

For a long time it was trusty established that surfactant molecules increase the dye solubility and interfere aggregation of dye molecules in water solutions. The reason of similar activity consists in formation and constitution direct (or inverse) micelles, whose exterior spherical (or cylindrical) parts possess the hydrophilic ability and the interior part of a micelle shows hydrophobic properties. For ionic dyes (R6G and majority dyes) the unorganic anion is hydrophilic while the responsible for optical parameters cation is partially hydrophobic. Both ions have been connected by force Coulomb interaction. Electrostatic energy minimization between dye and a solution will be reached, if the cation is localised in a hydrophobic kern of a micelle, and the anion settles down in a hydrophilic shell. In contribution /13/ this ability already was used for extruction of organic ion dye from a solution in SRSF on the base of ionic surfactant STAB. It was established that for effective extruction the sign of organic ion charge has prevailing value. Unlike STAB micelles with rather small kern, Pluronic P123 micelles have the major both leaky packed kern and the shell, therefore the similar micelle kern can to "hide" some dye organic cations. This property of micelles is termed as solulability, but amount of accepted inside molecules is evidently limited. Therefore since above mentioned concentrations of 0,5m/l is echieved the hydrophobic kern is saturated by cations and further they are forced to search the place on shell periphery that is to be an energy unsuitable and that is why the diffusion redistribution of dye concentration from homogeneous to nonhomogeneous with gradiant in a direction on periphery of a silicate film is taken place. Probably the ionic pairs with their characteristic absorption spectrum appear and are localized in the interface silicate wall-micelle shell.

For anion dyes the concentration dependence of spectrums in similar conditions will be definitely other and that object now is under study. Zwitterion dye PM580 forms more complicated 3 component spectrum in SRSF, than R6G and its detailed studying still will be coming (Fig. 7.).

The nonionic dyes are characterized by weak Van-der-Waals interaction that can not be observed in the ground state, while the interaction in vibronic excited state can appears itself as excimer/exciplex fluorescence.

A possibility of excimer fluorescence in complicated spectrums R6G at higher concentrations has been specified partially because the basic fluorescence band weakened and disappeared with concentration grows (fig. 4 and 5), while simultaneously the new band 2 with peak on 605-610nm appears and amplifies without the contour changes futher. For excimer formation it will be enough to localize at least 2 dye cations in a micelle kern, as for the known pirene excimer fluorescence originated at high concentrations even in an ordinary solution /14/.

Occurrence of a fluorescence emission with peak on 670нм and its approximate coincidence to the fluorescence band of a crystal R6G allows to connect it with minor reservations and care to the excitonic statets R6G in SRSF.

## THE INFERENCES

Evolution of absorption/fluorescence spectrums 3 types of dyes (with different nature of a charge state) at concentrations changes in structurally ranked silicate films with mesopores on a base of three-block copolymer Pluronic P123 is studied. It was found the monomolecular nature spectrum absorption/fluorescence are supported only up to definite concentration limit ($\sim$0,5m/l), but excess this limit is accompanied by the accruing changes.

The suggestion to the physical nature the observed spectral changes is formulated. It is related to amount of organic cations which can be localized in the hydrophobic micelle kern with their counterion in the hydrophilic micelle shell. At excess of the admissible «micelle» concentration the formation of the ionic dye pairs in the interface a silicate walls – micelle shell with the advent of an absorption spectrum of the ionic pair takes place. In absorption spectrums R6G in SRSF with dye concentration growth monotonously accrues the weight of absorption band of ionic pairs in a neighbourhood 500nm.

Dye localization in the interface is thermodynamically not suitable owing to the diffusion redistribution of the embeded dye with its transport on periphery of samples is observed.

At excess of amount of cation R6G in a hydrophobic kern more than 2 formation of excimers and excimer fluorescence with the characteristic spectrum and lifetime at the simultaneous supression of the known monomolecular fluorescent spectrum is observed. Accordingly, synchronous supression of the basic band weight R6G on 565nm is observed and the relative increase of band of excimer luminescence on 610nm and exciton emission on 670nm. On the possible exciton nature of emission R6G in SRSF at highest concentration specifies coincidence of given spectrum with R6G polycrystall spectrum.

## THE QUOTED LITERATURE

## FIGURE CAPTIONS

**Fig.1**. Diffractogram of SRSF without R6G (the upper dependence) and with R6G (the the bottom dependence).

**Fig.2.** Absorption spectrums R6G in SRSF (1) and polyurethanacrylate (2) at the equal optical densities, but the different concentrations (some orders of magnitude).

**Fig.3.** Evolution of absorption spectrum R6G with growth of concentrations in SRSF

**Fig.4a,b**) Evolution of spectrums of fluorescence R6G in SRSF with growth of concentration:1 - concentrations below 0,5M/l, 2-3-4- concentrations above the 0,5 M/l b) Deconvolution of integrated spectrum of fluorescence on 3 Gaussian contours

**Fig.5.** Fluorescence spectra R6G in SRSF at the various concentrations and in polycrystalline phase: contour 2 with a maximum on 605нм, contour 3-4 - limiting concentrations, prevails maximum 670нм, which is characteristic for fluorescence in polycrystalline phase R6G (5).

**Fig.6a,b.c.** Fluorescence lifetimes on various wavelengths inside integrated spectrums fig. 5.: a-contour 5, 670нм; b – contour 2, 582нм and 612нм; c- contour 4-630 and 670нм

**Fig.7**. Absorption spectrum PM580 in SRSF (1) and its deconvolution on 3 Gaussian form-factors (3,4,5) as the possible proof of multycenter existence, 2 – restitution of initial integrated contour with correlation factor COD (R^2) = 0,954

# FIGURES
**to paper "SPECTRUMS OF DYES IN THE MESOPOROUS SILICATE FILM"**

**FIG.1**.

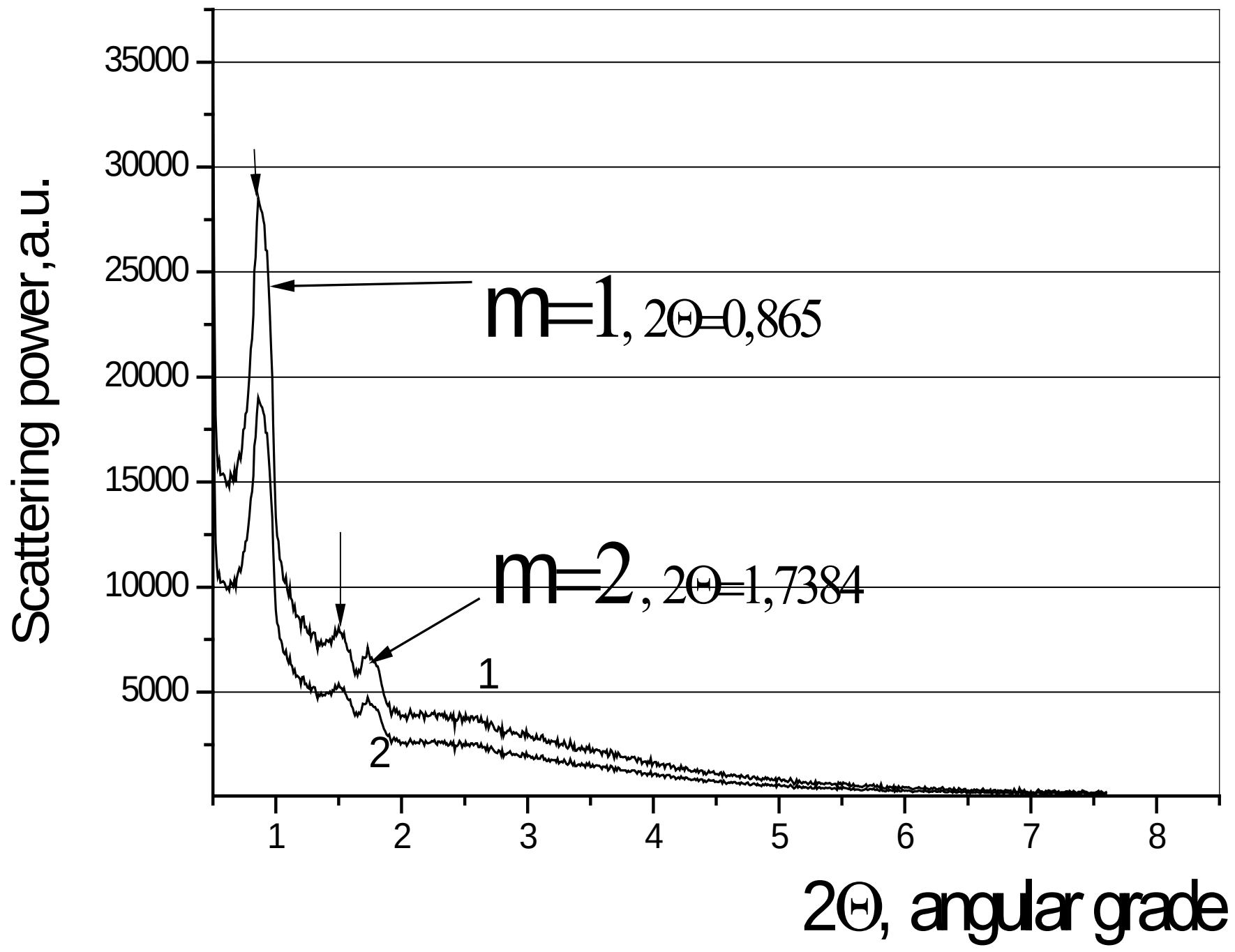

**FIG.2.**

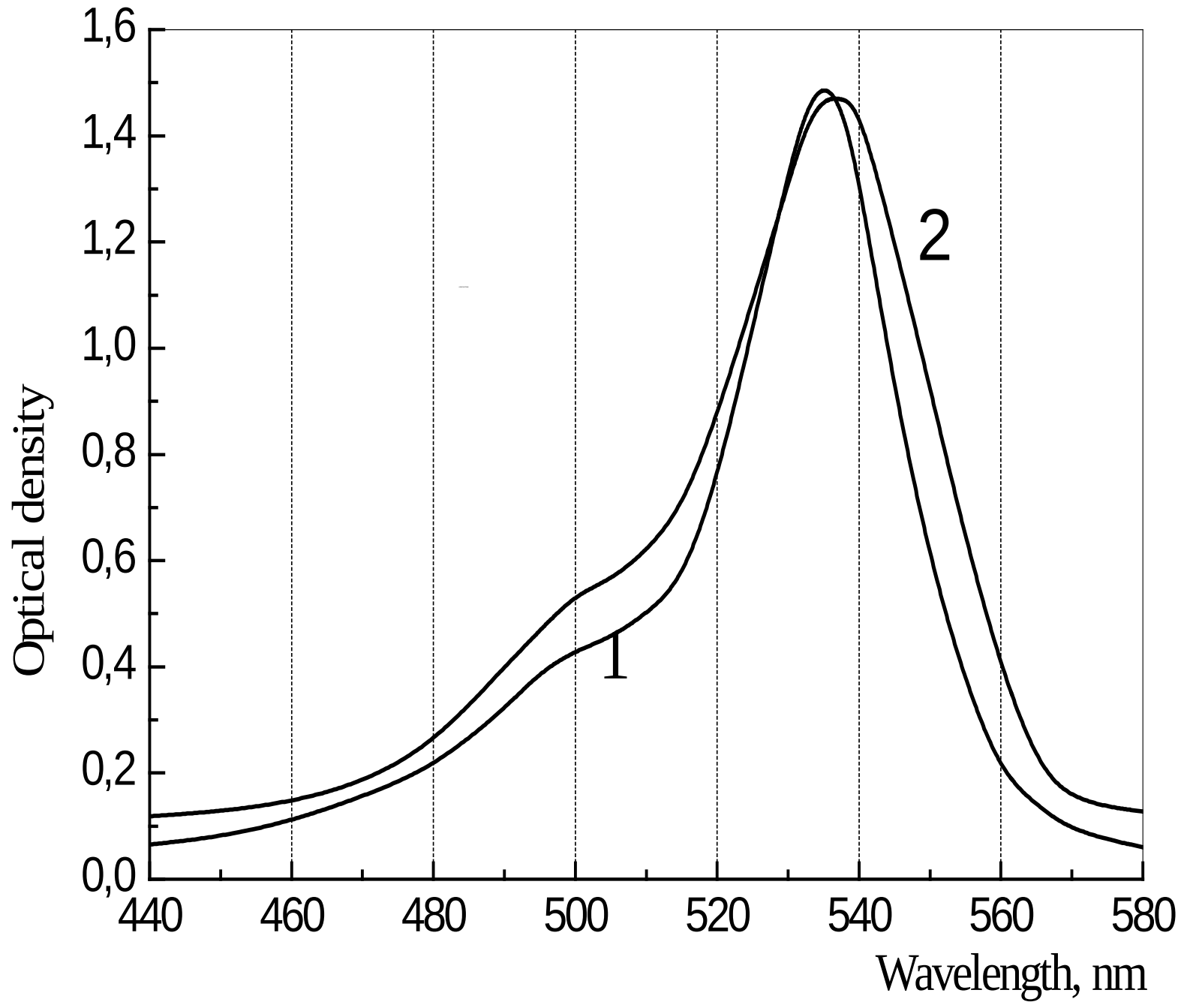

**To paper "SPECTRUMS OF DYES IN THE MESOPOROUS SILICATE FILM"**

**FIG.3.**

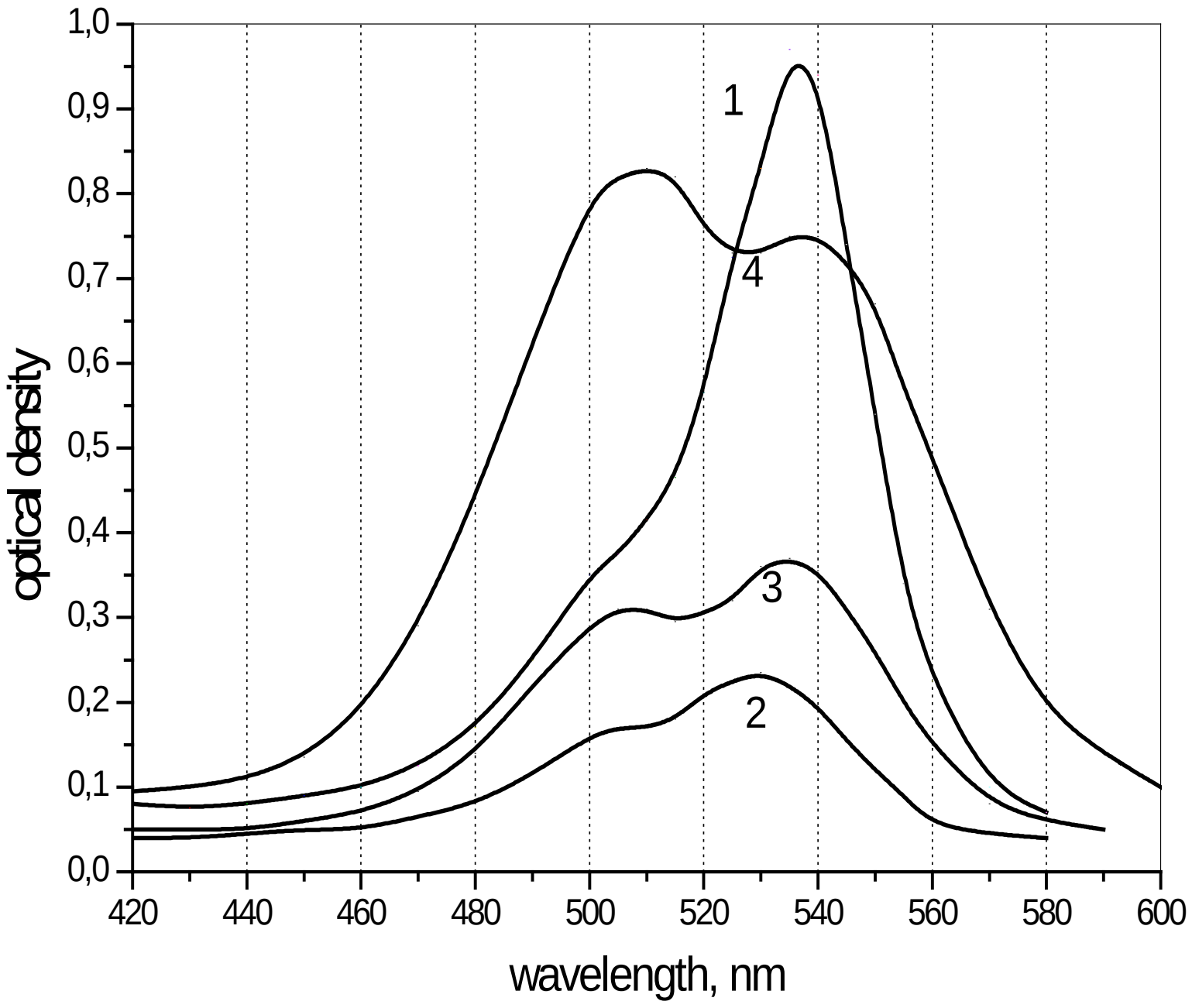

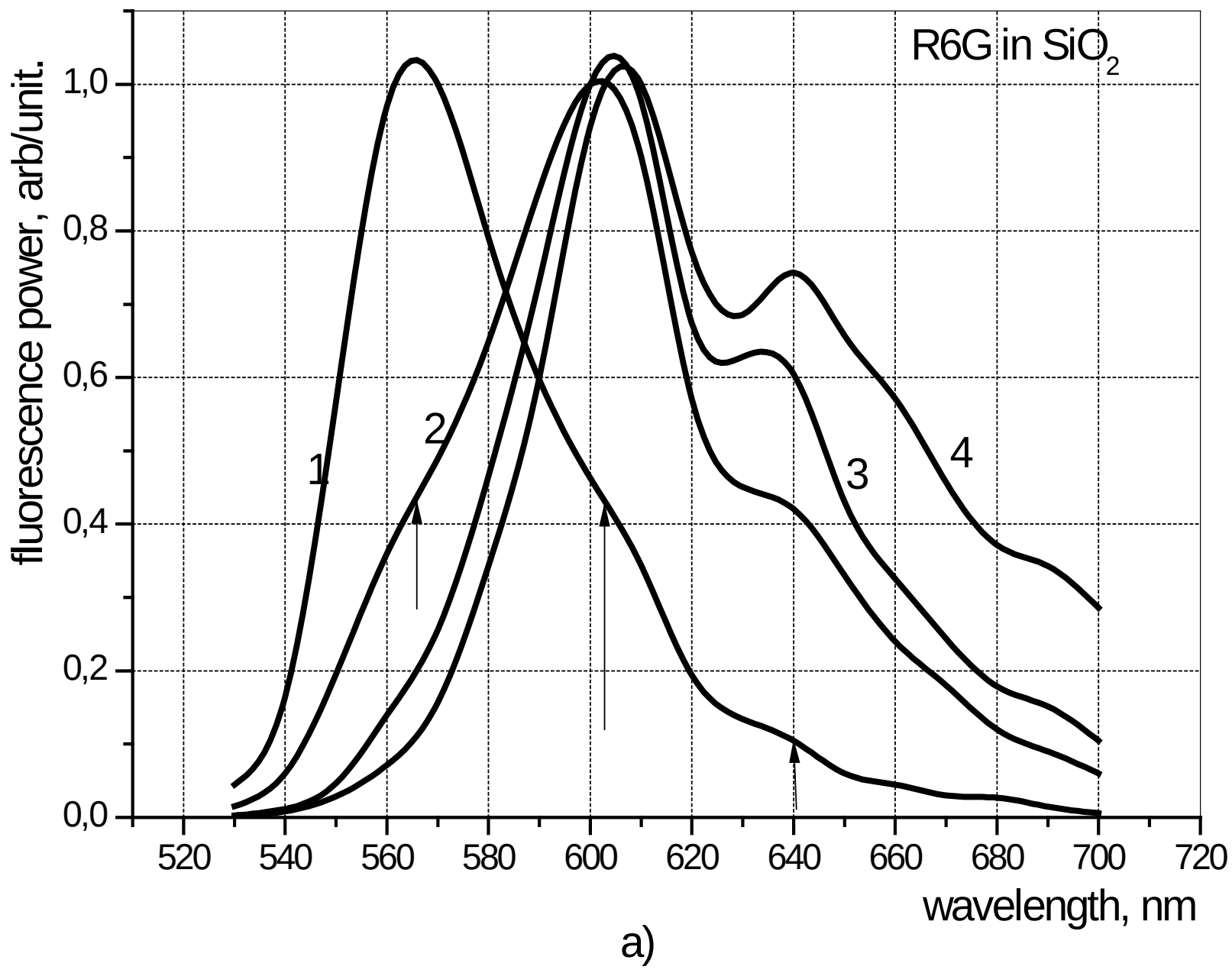

a)

**FIG.4a**

**to paper "SPECTRUMS OF DYES IN THE MESOPOROUS SILICATE FILM"**

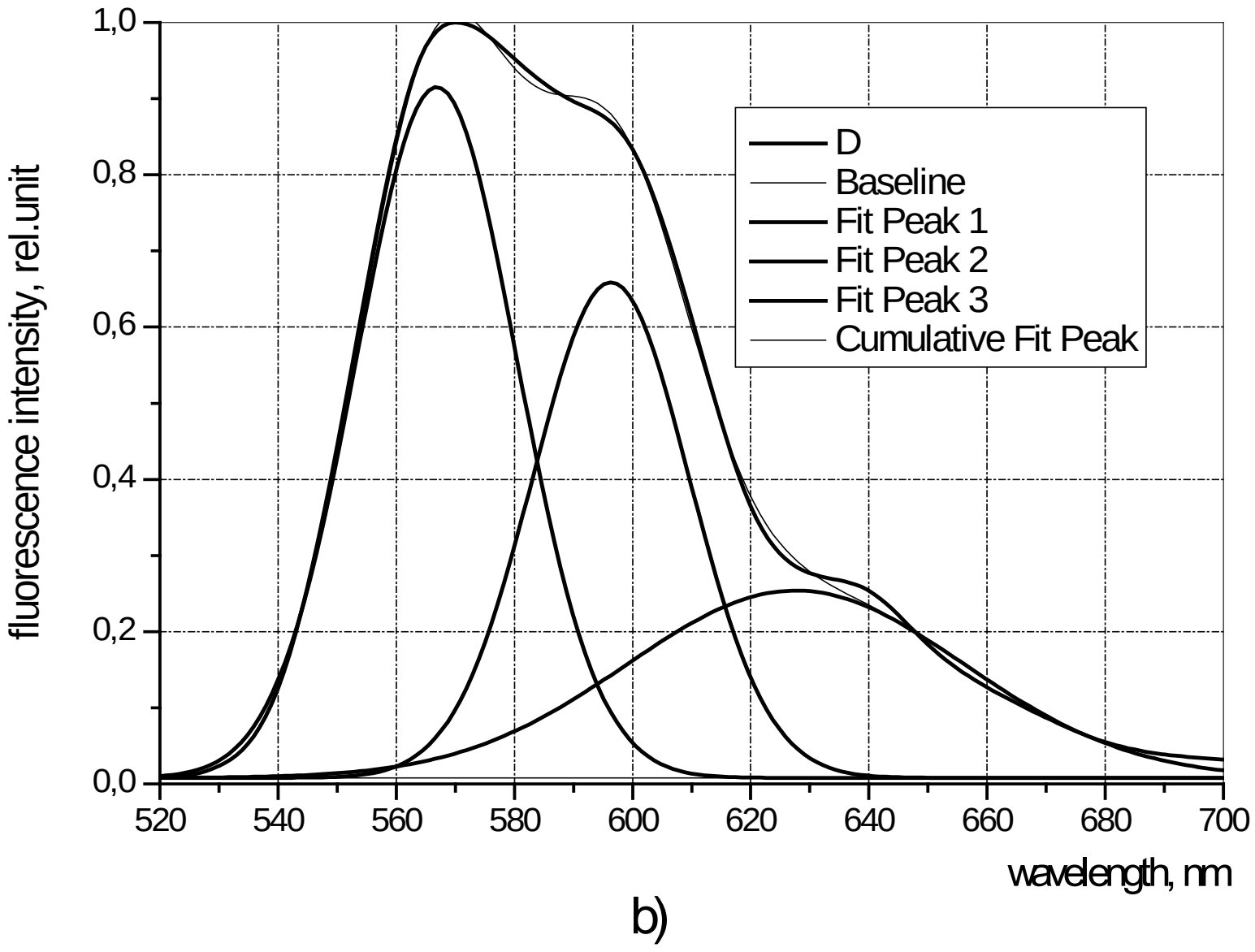

**FIG.4b**

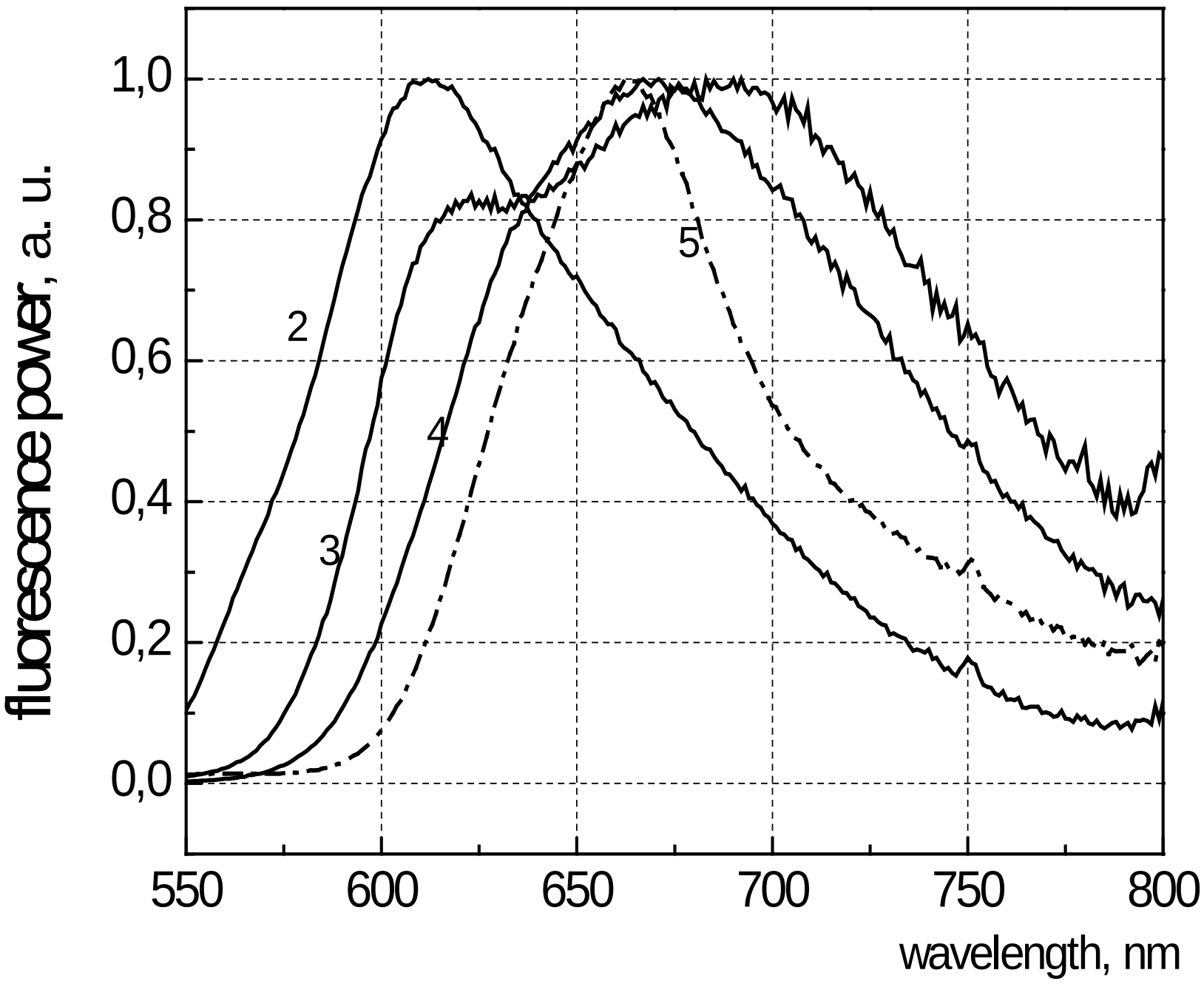

**FIG.5.**

**to paper "SPECTRUMS OF DYES IN THE MESOPOROUS SILICATE FILM"**

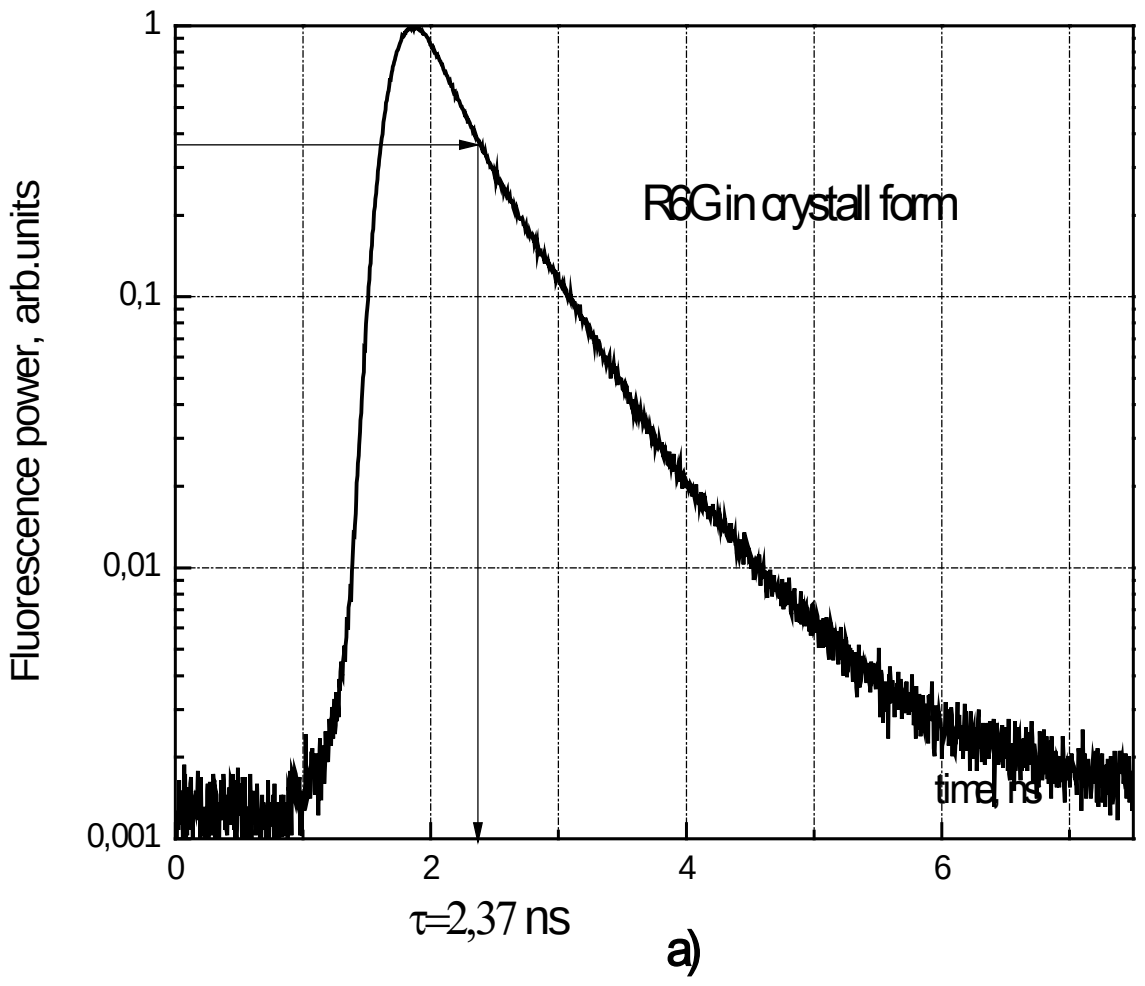

a)

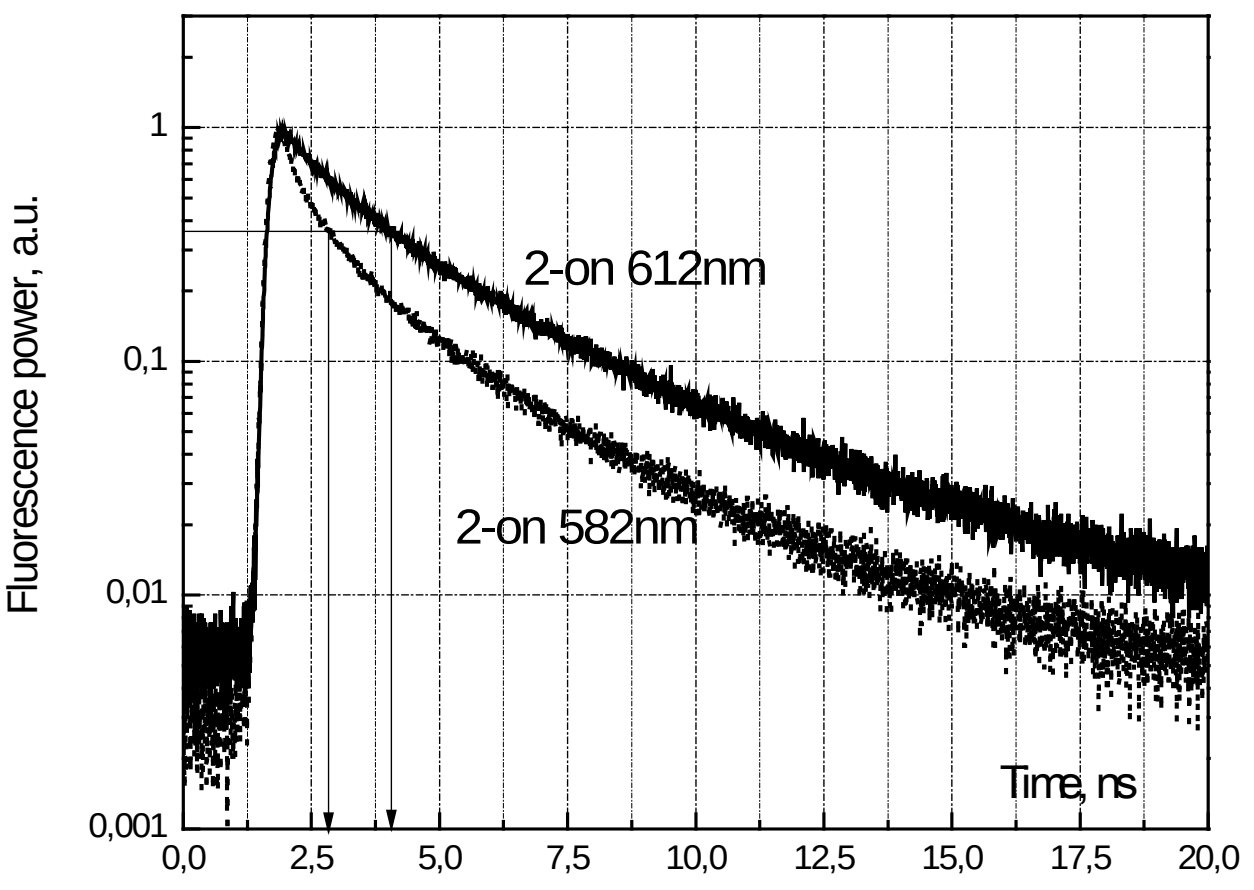

b)

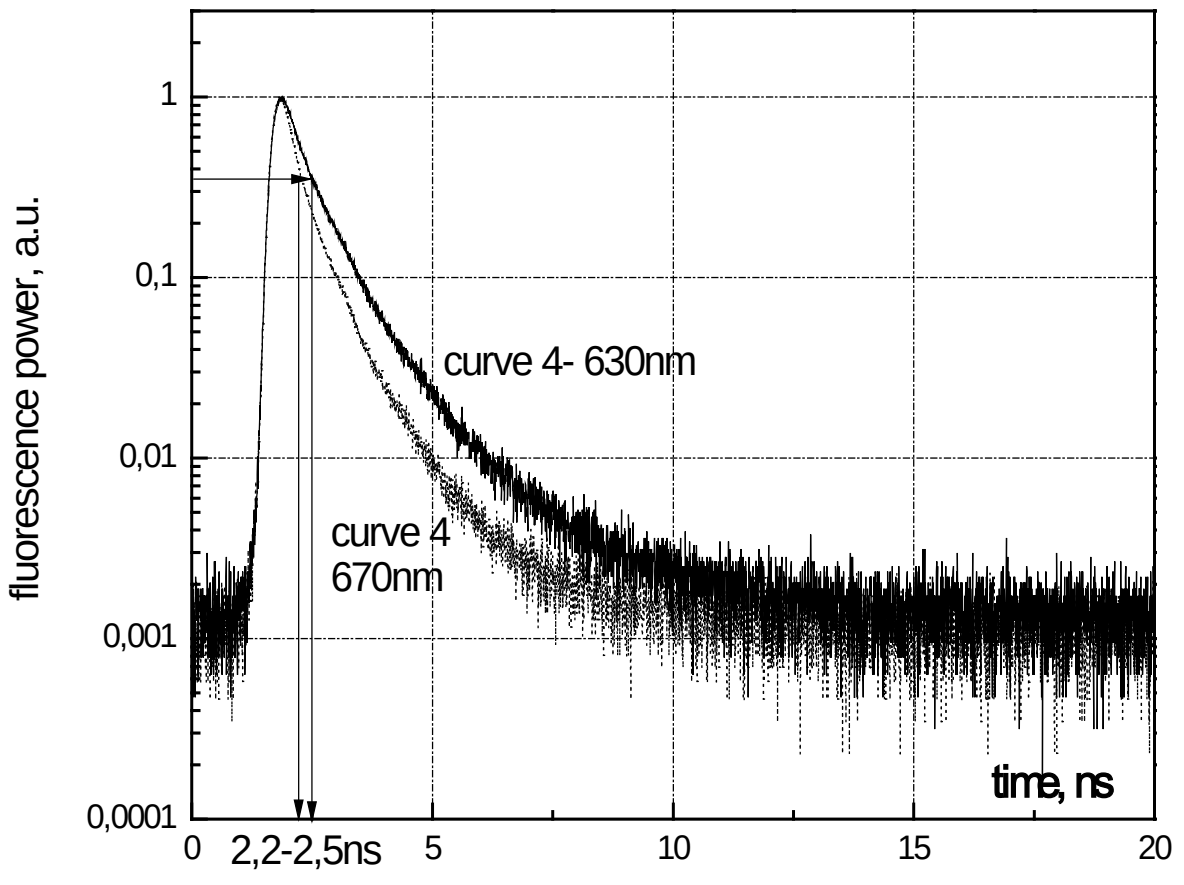

c)

**FIG.6a, b, c**

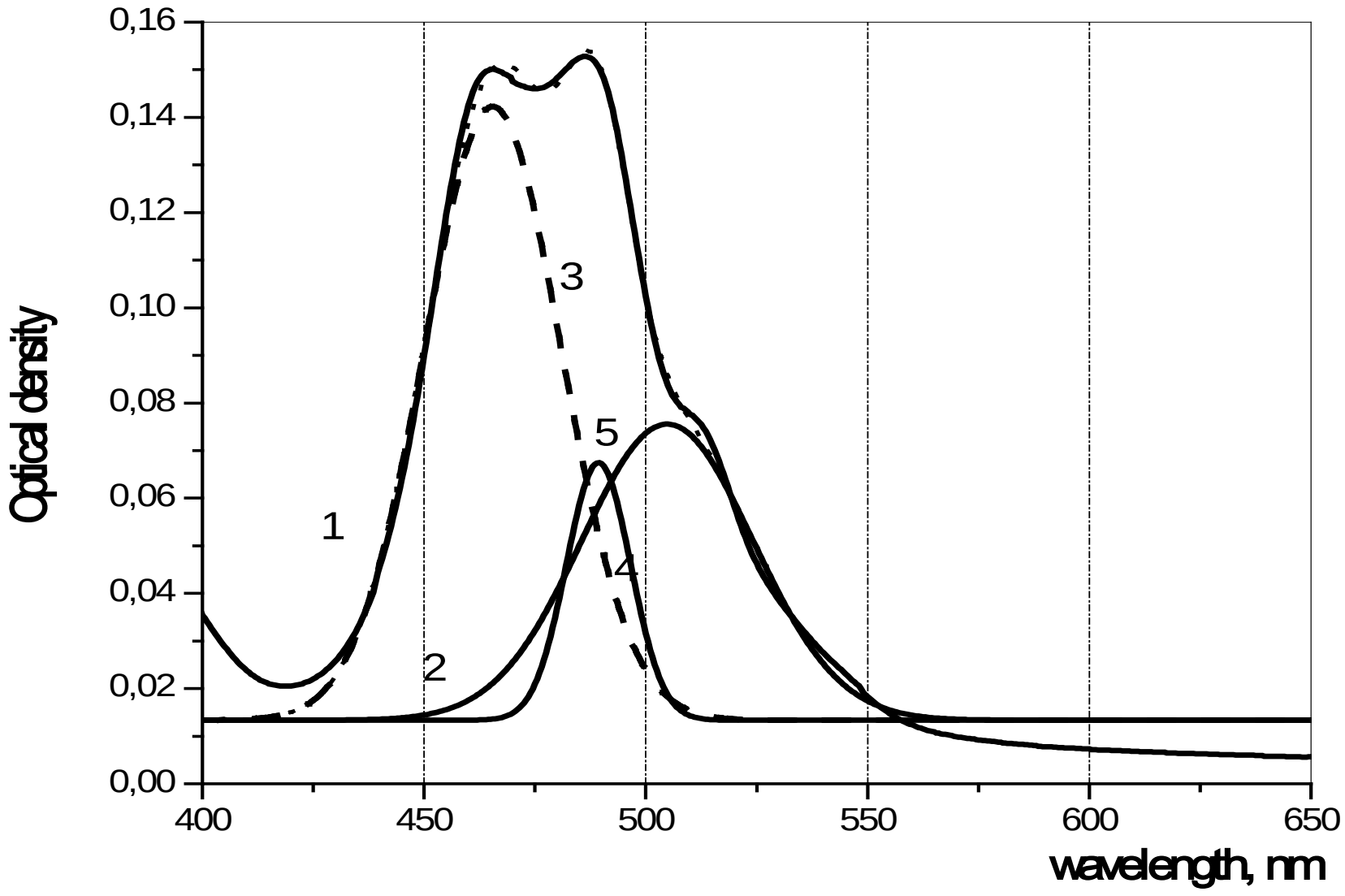

**FIG.7.**